\documentclass[aps,pre,twocolumn,groupedaddress,showpacs]{revtex4}

\usepackage{amsmath}
\usepackage{amssymb}
\usepackage{epsfig}
\newcommand {\dr}{{\mathrm d}\mathbf{r}}

\newcommand {\dd}{{\mathrm d}}
\newcommand {\rr}{\mathbf{r}}

\begin{document}

\title{Theory for the phase behaviour of a colloidal fluid with competing interactions}

\author{A.J. Archer$^1$}
\email{A.J.Archer@lboro.ac.uk}
\author{C. Ionescu$^{2,3}$}
\author{D. Pini$^2$}
\author{L. Reatto$^2$}
\affiliation{1. Department of Mathematical Sciences, Loughborough University, Loughborough, Leicestershire, LE11 3TU, United Kingdom\\
2. Dipartimento di Fisica, Universit\`a degli Studi di Milano, Via Celoria 16, 20133 Milano, Italy\\
3. Institute for Space Sciences P.O.Box MG-23, Ro 077125, Bucharest-Magurele, Romania}

\date{\today}

\begin{abstract}
We study the phase behaviour of a fluid composed of particles which interact via a pair potential that is repulsive for large inter-particle distances, is attractive at intermediate distances and is strongly repulsive at short distances (the particles have a hard core). As well as exhibiting gas-liquid phase separation, this system also exhibits phase transitions from the uniform fluid phases to modulated inhomogeneous fluid phases. Starting from a microscopic density functional theory, we develop an order parameter theory for the phase transition in order to examine in detail the phase behaviour. The amplitude of the density modulations is the order parameter in our theory. The theory predicts that the phase transition from the uniform to the modulated fluid phase can be either first order or second order (continuous). The phase diagram exhibits two tricritical points, joined to one another by the line of second order transitions.
\end{abstract}


\maketitle

\section{Introduction}
\label{sec:intro}

The interactions between colloidal particles can be complex and varied. In order to understand and determine the interactions one must not only consider the direct or `bare' interactions between the colloids themselves, but also the influence of the medium in which they are dispersed (the solvent). For example, if the colloids carry a net charge $q$, they will tend to repel each other. However, the strength of this repulsive interaction is determined not only by the magnitude of the charge $q$, but also by the concentration and types of counterions in the solvent. The counterions may condense on the surface of the colloids, forming an oppositely charged double-layer around the colloids, which screens the Coulomb interaction. Thus the range and strength of this repulsive interaction may be determined by controlling the concentration and type of coions and counterions in the solvent \cite{BarratHansen, Denton2007}.

The interaction between the colloids may also be influenced by the presence of other species in the solvent, such as polymeric macromolecules. If these polymers adhere to the surface of the colloids they form a soft layer surrounding the particles thus effectively increasing their size. Such adsorbing polymers may be used to stabilise the colloids in suspension. On the other hand, non-adsorbing polymers have the effect of generating an effective attraction between the colloids. This depletion interaction arises due to the fact that when the separation between two colloids becomes less than $\sim 2R_g$, where $R_g$ is the polymer radius of gyration, then the polymer chains are unlikely to be found between the two colloids  since such confinement entails an entropic cost. This depletion results in an unbalanced osmotic pressure on the colloids by the polymers, generating an effective attraction between the pair of colloids \cite{BarratHansen, Denton2007}.

In the last few years there have been a number of studies of systems of charged colloidal particles dispersed in a solvent containing nonadsorbing polymers \cite{Sedgwick1, Sedgwick2, Bartlett, Bartlett2}. Due to that fact that in these systems the charge is only weakly screened, there is a competition between a short ranged (depletion) attraction due to the presence of the polymers in the solution and a longer ranged (screened Coulomb) repulsive 
interaction. The competing interactions give rise to microphase separated fluid phases -- i.e.\ to fluid states exhibiting equilibrium modulated structures such as clusters or stripes. Other mechanisms leading to this kind of competition are also possible. For instance, the attraction may be due to dispersion forces instead of osmotic depletion. Even with neutral particles interacting solely via depletion forces, competition may arise as a consequence of mutual interactions between the depletants, which modify the interaction potential with respect to the simplest Asakura-Oosawa picture \cite{BarratHansen}. Structuring and pattern formation has also been seen in two dimensional systems \cite{GhezziEarnshawJPCM1997, SearetalPRE1999}, where the repulsion is likely to be due to dipole-dipole interactions between the particles at the surface. Together with these experimental studies, there has been significant theoretical interest in model colloidal systems exhibiting such competing interactions \cite{PinietalCPL2000, Pinietal2006, Archer19, Archer21, AndelamnetalJCP1987, KendricketalEPA1988, SearGelbartJCP1999, GroenewoldKegelJPCB2001, GroenewoldKegelJPCM2004, SciortinoetalPRL2004, MossaetalLangmuir2004, SciortinoetalJPCB2005, ImperioReattoJPCM2004, ImperioReattoJCP2006, ImperioReattoPRE2007, WuetalPRE2004, LiuetalJCP2005, BroccioetalJCP2006, Candiaetal2006, TarziaConiglioPRE2007, ConiglioetalJPCM2007, CharbonneauReichmanPRE2007, FalcuccietalEPL2008}. An overview of the results and conclusions drawn from this body of work is given in Ref.\ \cite{Archer21}, and will not be repeated here.

A simple model for these colloidal systems was considered by some of us previously \cite{PinietalCPL2000, Pinietal2006, Archer19, Archer21}. We model the `bare' interaction between the colloids as a hard-sphere interaction, the long ranged screened Coulomb repulsion with a Yukawa potential and we model the depletion attraction using an additional Yukawa potential. Thus the pair potential in our model is as follows:
\begin{equation}
v(r)=v_{hs}(r)+w(r)
\label{eq:v}
\end{equation}
where $v_{hs}(r)$ is the hard-sphere pair potential:
\begin{equation}
v_{hs}(r) = 
\begin{cases}
\infty \hspace{5mm} r \leq \sigma \\
0 \hspace{7mm} r > \sigma,
\end{cases}
\label{eq:v_hs}
\end{equation}
where $\sigma$ is the hard sphere-diameter and
\begin{equation}
w(r) = -\epsilon\sigma \frac{e^{- \lambda_1 \left(\frac{r}{\sigma}-1\right)}}{r}+A\sigma \frac{e^{- \lambda_2 \left(\frac{r}{\sigma}-1\right)}}{r}
\label{eq:w}
\end{equation}
The first term in $w(r)$ is the (attractive) depletion interaction and the second term is the (repulsive) screened Coulomb interaction. The magnitudes of the parameters $\epsilon$ and $A$ are determined by the concentration of non-adsorbing polymers in the solution and the charge carried by the colloids, respectively. The parameter $ \lambda_1  \sim \sigma/ R_g$ is roughly the size ratio between the colloids and the polymers and the parameter $ \lambda_2 $ is determined by the concentration and type of screening ions in the solvent. In the present work we consider the case when $ \lambda_1 > \lambda_2 $, so that the pair potential $v(r)$ is repulsive for large $r$ \cite{FortinietalJPCM2005}. However, at intermediate distances $r>\sigma$ between the particles, $v(r)$ may be attractive -- i.e.\ $v(\sigma^+)<0$. This occurs when the parameter $\epsilon>A$. In this case there is a competition between the long range repulsion and the shorter ranged attraction. Depending on the values of the set of pair potential parameters $\{ \epsilon,\, A,\,  \lambda_1 ,\,  \lambda_2 \}$, the fluid temperature $T$ and number density $\rho$, this model may exhibit various modulated phases \cite{Archer19, Archer21}.

The precise nature of the phase transition from the uniform fluid phase to the modulated fluid phases is not well understood. In Ref.\ \cite{Archer19} some of us developed a density functional theory (DFT) \cite{Bob, HM} for the present model. It was found that the DFT provided a good account of the structure and thermodynamics of the uniform liquid. The theory also predicts that for certain choices of the parameters $\{ \epsilon,\, A,\,  \lambda_1 ,\,  \lambda_2 \}$ and the temperature $T$, there exists an interval of the fluid density $\rho$ where the uniform fluid becomes unstable against periodic density fluctuations of a certain wavelength, indicating that there must be a phase transition to an inhomogeneous (modulated) phase. The locus of this line in the phase diagram is called the $\lambda$-line \cite{Archer19, StellJSP1995, CiachetalJCP2003, PatsahanCiachJPCM2007, Archer6}.

Following the study in Ref.\ \cite{Archer19}, a further study was made using Monte-Carlo (MC) computer simulations and integral equation theories \cite{Archer21}, in order to determine the nature of phase transition occurring at the $\lambda$-line. The results from the integral equation theories were mostly inconclusive -- all the closure relations that were implemented either failed to describe the phase transition or had no solution in the region of the phase diagram where the inhomogeneous phases are expected, with the exception of the Percus-Yevick (PY) closure, which is able to describe to some extent the nature of the transition from the low density homogeneous phase to the modulated (cluster) phase. However, this theory also failed to have a solution at higher densities.

The picture that emerged from the MC simulations was that for certain values of $A$ and $\epsilon^{-1}$ there is a first order phase transition from a low density homogeneous phase to a modulated (cluster) phase and also at higher densities a first order phase transition from a modulated (bubble) phase to the uniform liquid \cite{Archer21}. It was found that there is a difference in the densities $\Delta \rho$ between these coexisting phases. However, $\Delta \rho$ was found to be a fairly small quantity, which means that the density $\rho$ is not a good order parameter for this phase transition. We show below that the amplitude of the density modulations ${\cal A}$ proves to be a better order parameter for this phase transition. On following either of these two first order transitions to larger values of $\epsilon^{-1}$, one finds that $\Delta \rho \to 0$. The two critical points, at which these two transitions cease to be first order (i.e.\ when $\Delta \rho=0$) are difficult to locate using MC simulations \cite{Archer21}. One reason for this was due to a lack of knowledge of the relevant order parameter for the phase transition. Based on a comparison with  the phase behaviour of lattice models with competing interactions \cite{Barbosa93, Archer21}, it was concluded that the two critical points were most likely to be tricritical points \cite{ChaikinLubensky}, connected to one-another by a line of second order (continuous) transitions. However, it was not possible to see any signature of the second order transition line in the MC simulation results, leaving the conclusions somewhat tentative.

In the present work we use DFT to examine the phase behaviour in the vicinity of the $\lambda$-line and we find that the theory confirms the scenario proposed in Ref.\ \cite{Archer21} -- i.e.\ that the critical points are indeed tricritical points and that connecting these is the $\lambda$-line itself -- a line of second order transitions. We also use the DFT theory to develop an order parameter (Landau) theory for the phase transition, by making a sinusoidal approximation for the density modulations and then expanding the free energy in powers of the order parameter ${\cal A}$, allowing us to investigate the transition in detail. We also confirm our results numerically, solving the full DFT theory for the density profiles.

This paper proceeds as follows: In Sec.\ \ref{sec:DFT} we describe the simple mean field DFT that we use to study the phase transitions. In Sec.\ \ref{sec:fluid_structure} we determine structural properties predicted for the uniform fluid phase, focusing in particular on the static structure factor. We find a simple expression for the wave number at which the static structure factor has a peak, corresponding to the typical length scale of the fluid modulations. In Sec.\ \ref{sec:OP_theory}, starting from our DFT we develop an order parameter theory for the phase transitions in this system, by expanding the free energy in powers of ${\cal A}$. From our approximate free energy we determine the phase diagram of the system for particular choices of the pair potential parameters. In Sec.\ \ref{sec:DFT_res} we determine the phase behaviour by solving the full DFT, showing that our simplified order parameter theory captures most of the important physics. Finally, in Sec.\ \ref{sec:conc} we draw our conclusions.

\section{Mean-field DFT for the model}
\label{sec:DFT}

\begin{figure}
\includegraphics[width=1.\columnwidth]{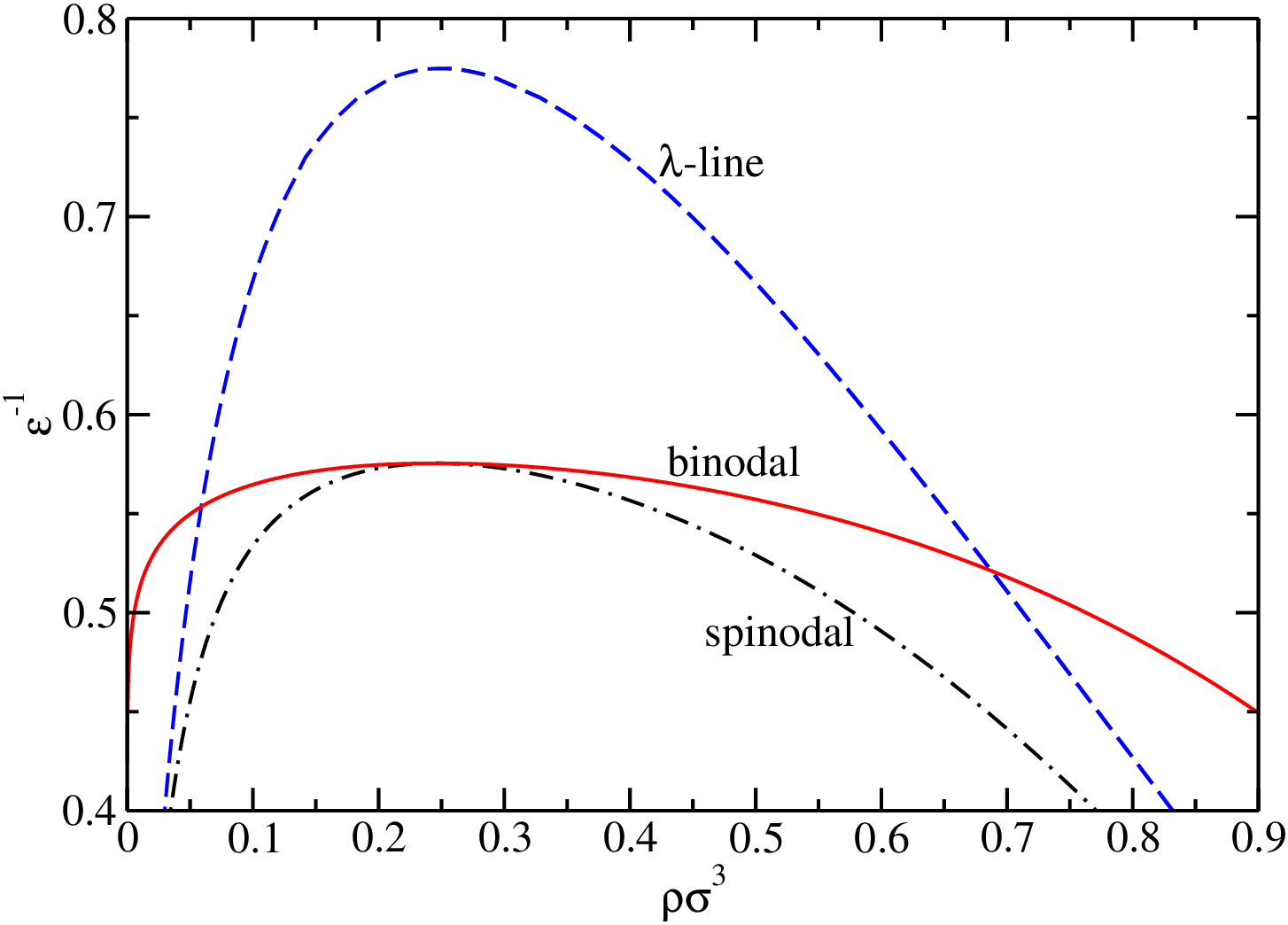}
\caption{Phase diagram in the $\epsilon^{-1}$ versus density $\rho$ plane obtained from the LDA DFT \eqref{eq:F} for the set of pair potential parameters $ \lambda_1 =1$, $ \lambda_2 =0.5$ and $A=0.5$. Within the region enclosed by the $\lambda$-line the uniform fluid is unstable against periodic density fluctuations with wavevector $k=k_c \neq 0$. The liquid-vapour coexistence curve (the binodal) and the spinodal are obtained from the same free energy functional for the uniform fluid. \label{lambdabase}}
\end{figure}

For systems of particles interacting via potentials of the form in Eq.\ \eqref{eq:v}, a common approach is to split the Helmholtz free energy into a contribution from the hard-sphere interactions between the particles -- the `reference' part -- and a separate part due to the slowly varying tail of the potential $w(r)$ \cite{HM}. Taking this approach, in Ref.\ \cite{Archer19}, the following approximation for the intrinsic Helmholtz free energy functional for the system was proposed:
\begin{eqnarray}
{\cal F}[\rho(\rr)]&=&{\cal F}_{hs}^{Ros}[\rho(\rr)] \notag \\
&+&\frac{1}{2} \int \dr \int \dr' \rho(\rr) \rho(\rr') w(\rr-\rr').
\label{eq:F_ros}
\end{eqnarray}
where  ${\cal F}_{hs}^{Ros}[\rho(\rr)]$ is the Rosenfeld approximation \cite{RosenfeldPRL1989, Rosenfeld:Levesque:WeisJCP1990, RosenfeldJCP1990} for the hard-sphere contribution to free energy. The Rosenfeld functional is a non-local weighted density functional, which also includes the exact ideal-gas contribution to the free energy. The remaining contribution is a simple mean-field approximation for the contribution to the free energy from the longer ranged interactions between the particles.

Note that in the mean field contribution to the free energy \eqref{eq:F_ros} [or \eqref{eq:F}], one requires the value of the potential $w(r)$ for all values of $r$. However, given the hard sphere contribution to the pair potential, Eqs.\ \eqref{eq:v} -- \eqref{eq:w}, the value of $w(r)$ for $0<r<\sigma$ should be irrelevant. Thus, the value of $w(r)$ for $0<r<\sigma$ used in the mean-field contribution to Eq.\ \eqref{eq:F_ros}, is effectively a free parameter in the theory. In Ref.\ \cite{Archer19} it was found that the choice $w(r)=w(\sigma^+)=-\epsilon+A$, for $0<r<\sigma$ gave better agreement with SCOZA for the fluid phase, than extending down to $0<r<\sigma$ the double Yukawa form of $w(r)$ that is used for $r>\sigma$. In the present study we use the truncated potential $w(r)=w(\sigma^+)=-\epsilon+A$, for $0<r<\sigma$ for all the DFT calculations. However, in Secs.\ \ref{sec:fluid_structure} and \ref{sec:OP_theory} we do not truncate $w(r)$ in order to simplify the analytic calculations.

The equilibrium one-body density profile $\rho(\rr)$ is determined by minimising the Grand potential functional
\begin{equation}
\Omega[\rho]={\cal F}[\rho] -\int \dr \rho(\rr)[\mu-V_{ext}(\rr)],
\label{eq:Omega}
\end{equation}
where $\mu$ is the chemical potential and $V_{ext}(\rr)$ is the external potential. The equilibrium density profile is the solution to the Euler-Lagrange equation
\begin{equation}
\frac{\delta \Omega[\rho]}{\delta \rho(\rr)} = 0.
\label{eq:min_Omega}
\end{equation}
If one requires the theory to be able to describe the oscillatory (with wavelength $\sim \sigma$) density profiles that occur when the fluid is subject to an external potential that varies strongly over short distances (such as, for example, the external potential due to the wall of the fluid container), then one must use a weighted-density DFT such as the Rosenfeld theory. However, if one is more interested in the large scale structures (stripes, clusters etc) that arise in the present system, due to the competing interactions in $w(r)$, then one may simplify the above theory, by making a local density approximation (LDA) in the reference hard sphere functional. In this case one may assume that the intrinsic Helmholtz free energy functional of the system is given by the following mean-field approximation:
\begin{eqnarray}
{\cal F}[\rho(\rr)]&=&\int \dr \rho(\rr)f(\rho(\rr))\notag \\
&+&\frac{1}{2} \int \dr \int \dr' \rho(\rr) \rho(\rr') w(\rr-\rr')
\label{eq:F}
\end{eqnarray}
where $f(\rho)$ is the Helmholtz free energy per particle of a uniform fluid of hard-spheres with bulk density $\rho$. We use the Carnahan-Starling approximation \cite{HM}:
\begin{equation}
\beta f(\rho)=\ln(\eta)+\frac{\eta(4-3\eta)}{(1-\eta)^2},
\label{eq:f_hd}
\end{equation}
where $\eta=\pi \rho \sigma^3/6$ is the packing fraction and $\beta=1/k_BT$ is the inverse temperature, which we will henceforth set to be $\beta=1$. From Eqs.\ \eqref{eq:min_Omega} -- \eqref{eq:f_hd} we obtain the following Euler-Lagrange equation for the equilibrium density profile:
\begin{equation}
f(\rho(\rr))+\rho(\rr)f'(\rho(\rr))+\int \dr' \rho(\rr') w(\rr-\rr')+V_{ext}(\rr)=\mu.
\label{eq:EL_eq_model}
\end{equation}

\begin{figure}
\includegraphics[width=1.\columnwidth]{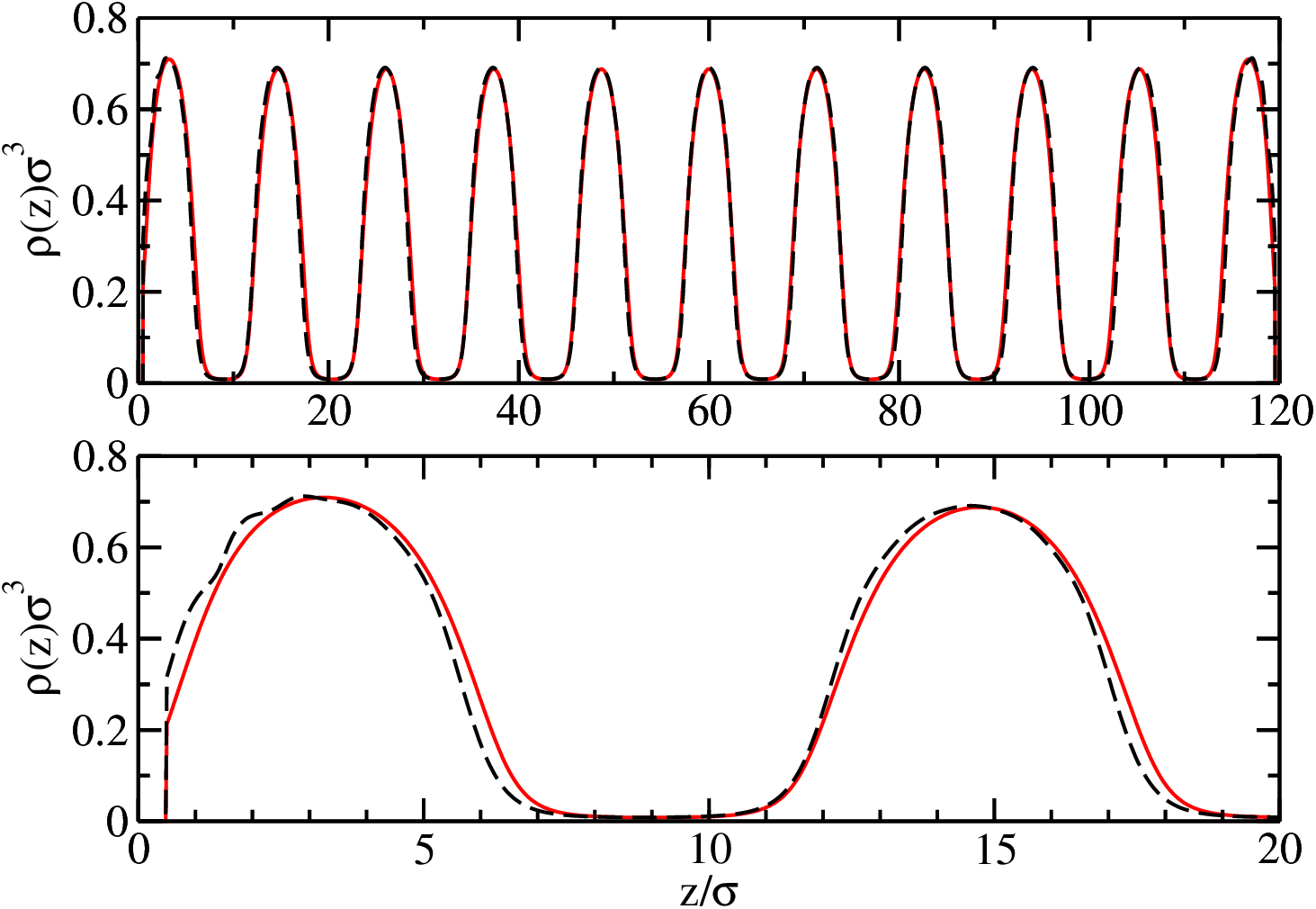}
\caption{\label{fig:slit_profiles} In the upper figure we display the density profile for the fluid confined between two parallel hard walls, separated a distance $L=120\sigma$, in the case when the pair potential parameters are $\epsilon^{-1}=0.65$, $A=0.5$, $ \lambda_1 =1$, $ \lambda_2 =0.5$ and the average fluid density is $\bar{\rho}\sigma^3=0.25$. In the lower figure, we display a magnification of the left hand portion of the density profile. The solid (red) line is the density profile calculated using the LDA reference hard sphere functional, Eq.\ \eqref{eq:F}. The dashed (black) line is the result obtained using the Rosenfeld functional for the hard-sphere reference functional.}
\end{figure}

The free energy of the uniform bulk liquid may be obtained by setting the density profile $\rho(\rr)=\rho$, i.e.\ a constant, in either of Eqs.\ \eqref{eq:F_ros} or \eqref{eq:F}. The resulting free energies differ slightly between the two functionals, since for a uniform fluid the hard-sphere contribution from the Rosenfeld functional \eqref{eq:F_ros} is equivalent to that from the scaled particle (PY compressibility) equation of state \cite{HM}, whereas the hard-sphere contribution to the free energy from \eqref{eq:F} is that from the Carnahan-Starling equation of state. However, for the densities of interest here, the difference between the two free energies is small. 

In Fig.\ \ref{lambdabase}, we display the uniform vapour-liquid coexistence curve (binodal), the spinodal and the $\lambda$-line, for the set of pair potential parameters: $ \lambda_1 =1$, $ \lambda_2 =0.5$ and $A=0.5$.
Here, as in Refs.~\cite{Archer19, Archer21}, the phase diagram has
been mapped by changing the attraction strength $\epsilon$ at fixed 
repulsion strength $A$. This can be achieved in a solution of charged 
colloidal particles by changing the depletant concentration 
at fixed temperature and salt concentration. This choice implies that the
region of $\epsilon$ values where competition is important and microphase
formation is expected has both a lower and an upper bound: at low
$\epsilon$, the interaction will be mostly or entirely repulsive, 
the $\lambda$-line is not met, 
and no transition will take place, except for those involving 
the occurrence of solid phases. At high $\epsilon$, the interaction will
be mostly attractive, and bulk liquid-vapour phase separation will take 
over. This is shown in Fig.\ \ref{lambdabase}, where the liquid-vapour 
binodal moves out of the region bounded by the $\lambda$-line 
for values of $\epsilon^{-1} \gtrsim 0.55$.

In order to demonstrate the reliability of the LDA DFT versus the more sophisticated Rosenfeld DFT, we calculate the fluid density profiles for the fluid confined between two parallel planar hard walls separated a distance $L=120 \sigma$, where $V_{ext}(\rr)=V_{ext}(z)=0$ for $0<z<L$ and $V_{ext}(z)=\infty$ otherwise. The external potential varies only in the Cartesian $z$-direction, and we assume that the fluid density profile also only varies in the $z$-direction, so that the Euler-Lagrange equation \eqref{eq:EL_eq_model} becomes
\begin{equation}
\mu=f(\rho(z))+\rho(z) f'(\rho(z))+\int \dd z' \rho(z')\phi(z-z')+V_{ext}(z),
\label{miusolve}
\end{equation}
where $\phi(z)=\int \dd x \int \dd y \, w(\rr)$. We solve this equation for the equilibrium density profile $\rho(z)$ by discretising the density profile and using a simple iterative numerical procedure. In Fig.\ \ref{fig:slit_profiles} we display results for the case when the pair potential parameters are chosen to be $\epsilon^{-1}=0.65$, $A=0.5$, $ \lambda_1 =1$, $ \lambda_2 =0.5$ and the chemical potential is chosen so that the average fluid density in the slit is $\bar{\rho}\sigma^3=0.25$. As one can see from Fig.\ \ref{lambdabase} this corresponds to a state point that is well inside the $\lambda$-line, where the fluid is strongly modulated. Comparing the two density profiles displayed in Fig.\ \ref{fig:slit_profiles}, we see that they are are fairly similar. The main difference occurs for values of $z$ that are close to the confining walls. The Rosenfeld functional, which gives a better account of the hard sphere correlations, predicts that the fluid profile contains oscillations with wavelength $\sim \sigma$ in the vicinity of the wall, due to packing of the spheres at the wall. However, the density profile obtained from LDA functional \eqref{eq:F} does not have these small length scale modulations and only exhibits the larger length scale modulations that arise due to the competing interactions in $w(r)$.

\section{Structure of the uniform fluid}
\label{sec:fluid_structure}

From previous studies of the present model fluid \cite{PinietalCPL2000, Pinietal2006, Archer19, Archer21, SearGelbartJCP1999, SciortinoetalPRL2004, SciortinoetalJPCB2005, ImperioReattoJPCM2004, ImperioReattoJCP2006, ImperioReattoPRE2007, WuetalPRE2004, LiuetalJCP2005, BroccioetalJCP2006}, we know that the static structure factor $S(k)$ plays an important role in characterising the microphase structuring displayed by the system. One finds that there is a large peak in $S(k)$ at $k=k_c \ll 2 \pi/\sigma$, where $l_c\equiv2 \pi/k_c$ is the length scale associated with the density modulations in the system. $S(k)$ is given by the following expression
\begin{equation}
S(k)=\frac{1}{1-\rho \hat{c}(k)},
\label{eq:S_of_k}
\end{equation}
where $\hat{c}(k)$ is the Fourier transform of $c(r)$, the bulk fluid (Ornstein-Zernike) direct pair correlation function \cite{HM}. This function may be obtained from the free energy functional via the following relation \cite{Bob, HM}:
\begin{equation}
c(\rr,\rr')=-\beta\frac{\delta^2({\cal F}[\rho(\rr)]-{\cal F}_{id}[\rho(\rr)])}{\delta \rho(\rr) \delta\rho(\rr')},
\label{eq:c_2}
\end{equation}
where
\begin{equation}
{\cal F}_{id}[\rho(\rr)]=k_BT \int \dr \rho(\rr)[\ln (\Lambda^3\rho(\rr))-1]
\label{eq:F_id}
\end{equation}
is the ideal-gas contribution to the free energy; $\Lambda$ being the thermal wavelength. 
For the homogeneous bulk fluid where $\rho(\rr)=\rho$, we find $c(\rr,\rr')=c(|\rr-\rr'|)=c(r)$. From Eqs.\ \eqref{eq:F} and \eqref{eq:c_2} we obtain the following approximation for the pair direct correlation function
\begin{eqnarray}
c(\rr,\rr')=&\hspace{-2mm}
-\beta\left[2f'(\rho(\rr))+\rho(\rr)f''(\rho(\rr))-\frac{k_BT}{\rho(\rr)}\right] \delta(\rr-\rr')\notag \\
&-\beta w(\rr-\rr'),
\label{eq:c_model}
\end{eqnarray}
where $f'$ and $f''$ are the first and second derivatives of $f$ with respect to $\rho$ and $\delta(\rr)$ is the Dirac delta function. Thus the Fourier transform of this quantity evaluated for the bulk fluid is
\begin{eqnarray}
\hat{c}(k)=-\beta\left[2f'(\rho)+\rho f''(\rho)-\frac{k_BT}{\rho}\right]-\beta \hat{w}(k),
\label{eq:c_hat_model}
\end{eqnarray}
where $\hat{w}(k)$ is the Fourier transform of $w(r)$ and the first term in the right hand side is equal to $-1/(\rho \chi_{\it hs}^{\it red})+1/\rho$, $\chi_{\it hs}^{\it red}$ being the isothermal compressibility of the hard-sphere fluid divided by that of the ideal gas. This treatment amounts to taking the expression
\begin{eqnarray}
\hat{c}(k)=\hat{c}_{\it hs}(k)-\beta \hat{w}(k)
\label{eq:c_RPA}
\end{eqnarray}
given by the random phase approximation (RPA) and setting $k=0$ in $\hat{c}_{\it hs}(k)$, the Fourier transform of the hard-sphere direct pair correlation finction. In Ref.~\cite{Archer19} the uniform phase was studied by the full RPA, which is obtained from the free energy functional~(\ref{eq:F_ros}) adopted there in the homogeneous regime. If we define the following parameters: $\epsilon_1=\epsilon \exp( \lambda_1 )$ and $\epsilon_2=A\exp( \lambda_2 )$, then we may rewrite the pair potential $w(r)$ as follows:
\begin{equation}
w(r) = -\epsilon_1 \frac{ \exp[- \lambda_1 (r/\sigma)]}{r/\sigma} + \epsilon_2  \frac{ \exp[- \lambda_2 (r/\sigma)]}{r/\sigma}.
\label{eq:w_2}
\end{equation}
From this we obtain:
\begin{equation}
\hat{w}(k)=-\frac{4 \pi \epsilon_1 \sigma^3}{( \lambda_1 ^2+k^2\sigma^2)}
+\frac{4 \pi \epsilon_2 \sigma^3}{( \lambda_2 ^2+k^2\sigma^2)}.
\label{eq:w_hat}
\end{equation}
Thus the static structure factor is given by
\begin{eqnarray}
S(k)&=&\frac{1}{\rho \beta [2f'(\rho)+\rho f''(\rho)+ \hat{w}(k)]}\label{eq:S_of_k_model} \\
&\equiv& \frac{1}{D(k)},
\label{eq:D}
\end{eqnarray}
which defines the denominator function $D(k)$. For the sets of parameters $\{\lambda_1,\lambda_2,\epsilon_1,\epsilon_2\}$ that we consider here, the structure factor given by Eq.\ \eqref{eq:S_of_k_model} exhibits a single peak, at $k=k_c$. This is the peak that characterises the cluster/stripe modulated structures in the system. This approximation for $S(k)$ becomes unreliable for large wave numbers $k\gg k_c$, as one should expect, given the delta-function approximation for the hard-sphere contribution to $c(r)$, in Eq.\ \eqref{eq:c_model}.

One finds that in a certain portion of the phase diagram, the uniform fluid is unstable with respect to periodic density fluctuations -- this occurs when $S(k) \to \infty$ at $k=k_c$. In other words, when $D(k=k_c) \to 0$. The locus in the phase diagram at which $D(k=k_c)=0$ is the $\lambda$-line \cite{Archer19}. In Fig.\ \ref{lambdabase}, we display the $\lambda$-line for the set of pair potential parameters: $ \lambda_1 =1$, $ \lambda_2 =0.5$ and $A=0.5$. We see that this instability pre-empts the spinodal instability which corresponds to $S(k=0) \to \infty$; i.e.\ when $D(k=0) \to 0$.

The wavelength of the periodic density modulations is $l_c \equiv 2 \pi/k_c$; this length scale is the typical distance from the peak of one density modulation to the next. Within the present theory, we are able to obtain a simple expression for $k_c$, as a function of the pair potential parameters. To do this we recall that $k_c$ is the value of $k$ for which $S(k)$ is maximum. This maximum in $S(k)$ corresponds to a minimum in $D(k)$, i.e.\ when
\begin{eqnarray}
\frac{\partial D(k)}{\partial k}=0.
\end{eqnarray}
Given that within our approximation for $D(k)$, Eqs.\ \eqref{eq:S_of_k_model} and \eqref{eq:D}, the only $k$-dependence in $D(k)$ enters in the function $ \hat{w}(k)$, we find that this condition simplifies to the following:
\begin{eqnarray}
\frac{\partial \hat{w}(k)}{\partial k}=\frac{8 \pi \epsilon_1 \sigma^5 k }{( \lambda_1 ^2+k^2\sigma^2)^2}- \frac{8 \pi \epsilon_2 \sigma^5 k }{( \lambda_2 ^2+k^2\sigma^2)^2}=0.
\label{eq:k_c_eq}
\end{eqnarray}
There are two solutions to Eq.\ \eqref{eq:k_c_eq}. The first  is $k=0$. This corresponds to a maximum in $D(k)$ and a minimum in $S(k)$. The second solution to Eq.\ \eqref{eq:k_c_eq} is $k=k_c$, where
\begin{equation}
k_c=\frac{1}{\sigma}\sqrt{\frac{ \lambda_1 ^2-\alpha  \lambda_2 ^2}{\alpha-1}},
\label{eq:k_c}
\end{equation}
and where $\alpha=\sqrt{\epsilon_1/\epsilon_2}$. Note that this expression for $k_c$ does not depend of the fluid density $\rho$, indicating that the wavelength of the density modulations is independent of the fluid density. However, the amplitude of any density modulations {\em does} depend on the fluid density, since this quantity depends upon the height of the maximum in $S(k)$, which does depend on the density $\rho$.

\begin{figure}
\includegraphics[width=1.\columnwidth]{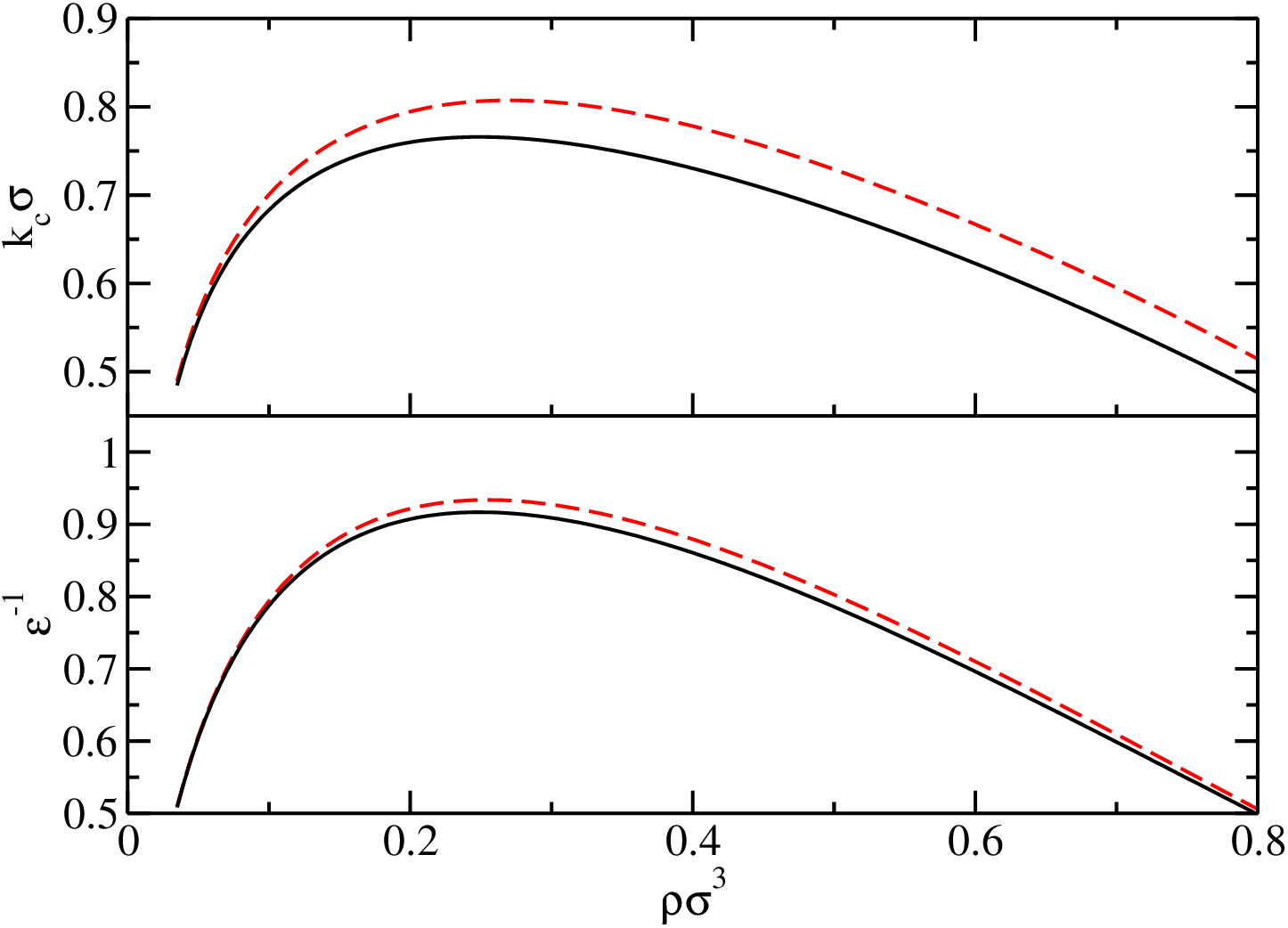}
\caption{\label{fig:lambda} In the upper figure we display $k_c$, the wave-number for the maximum in $S(k)$, calculated along the $\lambda$-line, for the case when $A=0.5$, $\lambda_1=1$ and $\lambda_2=0.5$. The solid line is the result in Eq.\ \eqref{eq:k_c} and the dashed line is the result from the RPA. Below this we display the $\lambda$-line. The solid line is the result in Eq.\ \eqref{eq:lambda_line_3} and the dashed line is the result from solving the RPA.}
\end{figure}

We are now in the position to obtain a relatively simple expression for the value of $\epsilon_1$ on the $\lambda$-line, as a function of density and the other pair potential parameters. Recall that the $\lambda$-line corresponds to the line in the phase diagram where $D(k=k_c)=0$ \cite{Archer19}. From Eqs.\ \eqref{eq:S_of_k_model} and \eqref{eq:D} we see that this condition is equivalent to the condition
\begin{eqnarray}
2f'(\rho)+\rho f''(\rho)+\hat{w}(k_c)=0.
\label{eq:lambda_line_1}
\end{eqnarray}
Using our expression for $k_c$ in Eq.\ \eqref{eq:k_c}, we obtain:
\begin{eqnarray}
\hat{w}(k_c)=-\frac{4 \pi \sigma^3\epsilon_2 (\alpha-1)^2}{\lambda_1 ^2- \lambda_2 ^2}.
\label{eq:lambda_line_2}
\end{eqnarray}
Eqs.\ \eqref{eq:lambda_line_1} and \eqref{eq:lambda_line_2} may be solved in a straightforward manner, by fixing the value of the density $\rho$ and then solving Eq.\ \eqref{eq:lambda_line_1} for $\epsilon_1$, to obtain:
\begin{eqnarray}
\epsilon_1=\epsilon_2\left( 1+\sqrt{\frac{(\lambda_1 ^2- \lambda_2 ^2)(2f'(\rho)+\rho f''(\rho))}{4 \pi \sigma^3 \epsilon_2}} \right)^2.
\label{eq:lambda_line_3}
\end{eqnarray}
This gives the value of $\epsilon_1$ on the $\lambda$-line, as a function of the fluid density, $\rho$.

In the lower panel of Fig.\ \ref{fig:lambda} we compare the result for the $\lambda$-line in Eq.\ \eqref{eq:lambda_line_3} with the result obtained from the RPA, Eqs.\ \eqref{eq:S_of_k} and \eqref{eq:c_RPA}. Recall that Eq.\ \eqref{eq:lambda_line_3} is obtained by taking the RPA expression for the pair direct correlation function, Eq.\ \eqref{eq:c_RPA}, and setting $k=0$ in $\hat{c}_{hs}(k)$. In the upper panel of Fig.\ \ref{fig:lambda} we display the value of $k_c$ calculated along the $\lambda$-line, obtained from Eq.\ \eqref{eq:k_c} (solid line) and also the result from the RPA, Eqs.\ \eqref{eq:S_of_k} and \eqref{eq:c_RPA}, (dashed line). We see fairly good agreement between the two sets of results, providing further confirmation that the the LDA free energy functional \eqref{eq:F} is able to account for the correlations in the fluid in the portion of the phase diagram in the vicinity of the $\lambda$-line.

\section{Order parameter theory}
\label{sec:OP_theory}

In this section, starting from the DFT proposed in Sec.\ \ref{sec:DFT}, we develop an order parameter theory for the phase transition from the uniform fluid to the modulated fluid phase. To develop the theory, we make the following assumptions: (i) we assume that the density profile varies only in one spatial direction, i.e.\ we assume $\rho(\rr) \to \rho(z)$. (ii) We assume the density modulations $\delta \rho(z)$  are of the form
\begin{eqnarray}
\rho(z)&=&\bar{\rho} +\delta \rho(z) \notag \\
&=& \bar{\rho} +{\cal A} \sin(k z),
\label{eq:OP}
\end{eqnarray} 
where $\bar{\rho}$ is the average fluid density and the amplitude ${\cal A}$ is the order parameter for the transition. We should expect the wave number for the modulations in Eq.\ \eqref{eq:OP} to be $k=k_c$. However, for the present we will not make any assumptions concerning the precise value of $k$ in Eq.\ \eqref{eq:OP}, other than to assume it is finite and non-zero.

From our assumption that the density profile varies in only one Cartesian direction we may re-write the Helmholtz free energy \eqref{eq:F} as follows:
\begin{eqnarray}
{\cal F}[\rho(z)]&=&L^2 \int \dd z \rho(z)f(\rho(z))\notag \\
&+&\frac{L^2}{2} \int \dd z \int \dd z' \rho(z) \rho(z') \phi(z-z')
\label{eq:F_z}
\end{eqnarray}
where the limits on the integrals go from $-L/2$ to $L/2$ (we implicitly take the thermodynamic limit $L \to \infty$), $L^2=\int \dd x \int \dd y$ is the system cross sectional area and
\begin{eqnarray}
\phi(z)&=&\int \dd x \int \dd y \, w(\rr) \notag \\
&=&-\frac{2 \pi \sigma^2 \epsilon_1}{ \lambda_1 } e^{- \lambda_1 z/\sigma}+\frac{2 \pi \sigma^2 \epsilon_2}{ \lambda_2 } e^{- \lambda_2  z/\sigma}
\label{eq:int_w}
\end{eqnarray}
where we have used Eq.\ \eqref{eq:w_2} to obtain the second line in Eq.\ \eqref{eq:int_w}.

Substituting Eq.\ \eqref{eq:OP} into \eqref{eq:F_z} and making a Taylor expansion in the LDA hard-sphere part of the free energy, we obtain:
\begin{eqnarray}
{\cal F}[\rho]  &=& L^2 \int \dd z \Bigg\{ \bar{\rho} f(\bar{\rho})
+ \frac{1}{2} \frac{\partial^2 (\rho f)}{\partial \rho^2}\Bigg\vert_{\bar{\rho}} \delta \rho^2 \notag \\
&\,&+ \frac{1}{4!} \frac{\partial^4 (\rho f)}{\partial \rho^4}\Bigg\vert_{\bar{\rho}} \delta \rho^4
+\frac{1}{2}\bar{\rho}^2 \hat{w}(0)+\frac{1}{2} \hat{w}(k)\delta\rho^2 \notag \\
&\,&+{\cal O}(\delta \rho^6) \Bigg\}\\
&=& {\cal F}[\bar{\rho}] +\Bigg\{ \frac{\partial^2 (\rho f)}{\partial \rho^2} \Bigg\vert_{\bar{\rho}} +\hat{w}(k) \Bigg\} \frac{L^2}{2} \int \dd z \, \delta \rho^2 \notag \\
&\,&+ \frac{\partial^4 (\rho f)}{\partial \rho^4}\Bigg\vert_{\bar{\rho}}\frac{L^2}{4!}  \int \dd z \, \delta \rho^4+{\cal O}( \int \dd z \, \delta \rho^6)
\end{eqnarray}
where we have used the fact that all terms which involve an integral over an odd power of $\delta \rho$ are zero, due to the fact that $\delta \rho$ is a sinusoidal function of $z$. If we now assume that the system size $L=\pi m /k$, where $m$ is an integer, then we obtain:
\begin{eqnarray}
\frac{\beta {\cal F}[\rho]}{V}=\frac{\beta {\cal F}[\bar{\rho}]}{V}+\frac{D(k)}{4\bar{\rho}} {\cal A}^2+\frac{B}{64} {\cal A}^4+{\cal O}({\cal A}^6)
\label{eq:Landau}
\end{eqnarray}
where $V=L^3$ is the system volume,
\begin{eqnarray}
\frac{\beta {\cal F}[\bar{\rho}]}{V}=\bar{\rho}f(\bar{\rho})+\frac{1}{2}\bar{\rho}^2\hat{w}(0)
\label{eq:F_bulk}
\end{eqnarray}
is the free energy of the uniform fluid and $B=\beta [\partial^4 (\rho f)/\partial \rho^4 ]_{\bar{\rho}}=\beta(4f'''(\bar{\rho})+\bar{\rho}f''''(\bar{\rho}))$. The higher order terms in the expansion \eqref{eq:Landau} are fairly straightforward to obtain, since they are merely terms arising from the Taylor expansion of the hard-sphere part of the free energy functional. For example, the next term of order ${\cal O}({\cal A}^6)$ in Eq.\ \eqref{eq:Landau}, is $C{\cal A}^6/6$, where $C=\beta [\partial^6 (\rho f)/\partial \rho^6 ]_{\bar{\rho}}/384$.

We may now use Eq.\ \eqref{eq:Landau} to investigate the phase behaviour in the vicinity of the $\lambda$-line. For a given state point, the equilibrium value of the amplitude ${\cal A}$ is that which minimises the free energy \eqref{eq:Landau} -- i.e.\ the equilibrium value of the amplitude is the solution to the equation
\begin{eqnarray}
\frac{\partial {\cal F}}{\partial {\cal A}}=0,
\label{eq:min_cond}
\end{eqnarray}
which together with \eqref{eq:Landau} gives us the following equation to be solved for ${\cal A}$:
\begin{eqnarray}
\frac{D(k)}{2\bar{\rho}} {\cal A}+\frac{B}{16} {\cal A}^3+{\cal O}({\cal A}^5)=0
\label{eq:min_cond_2}
\end{eqnarray}
We first note that the coefficient $B>0$ for all densities $0<\bar{\rho}<6/\pi \sigma^3$ and the same is true for all coefficients of higher order terms in the expansion. Secondly, we note that for the uniform fluid, outside the $\lambda$-line, $D(k)>0$ for all values of $k$. Thus, outside the $\lambda$-line, the minimum of the free energy \eqref{eq:Landau} is when ${\cal A}=0$, as one should expect.

\begin{figure}
\includegraphics[width=1.\columnwidth]{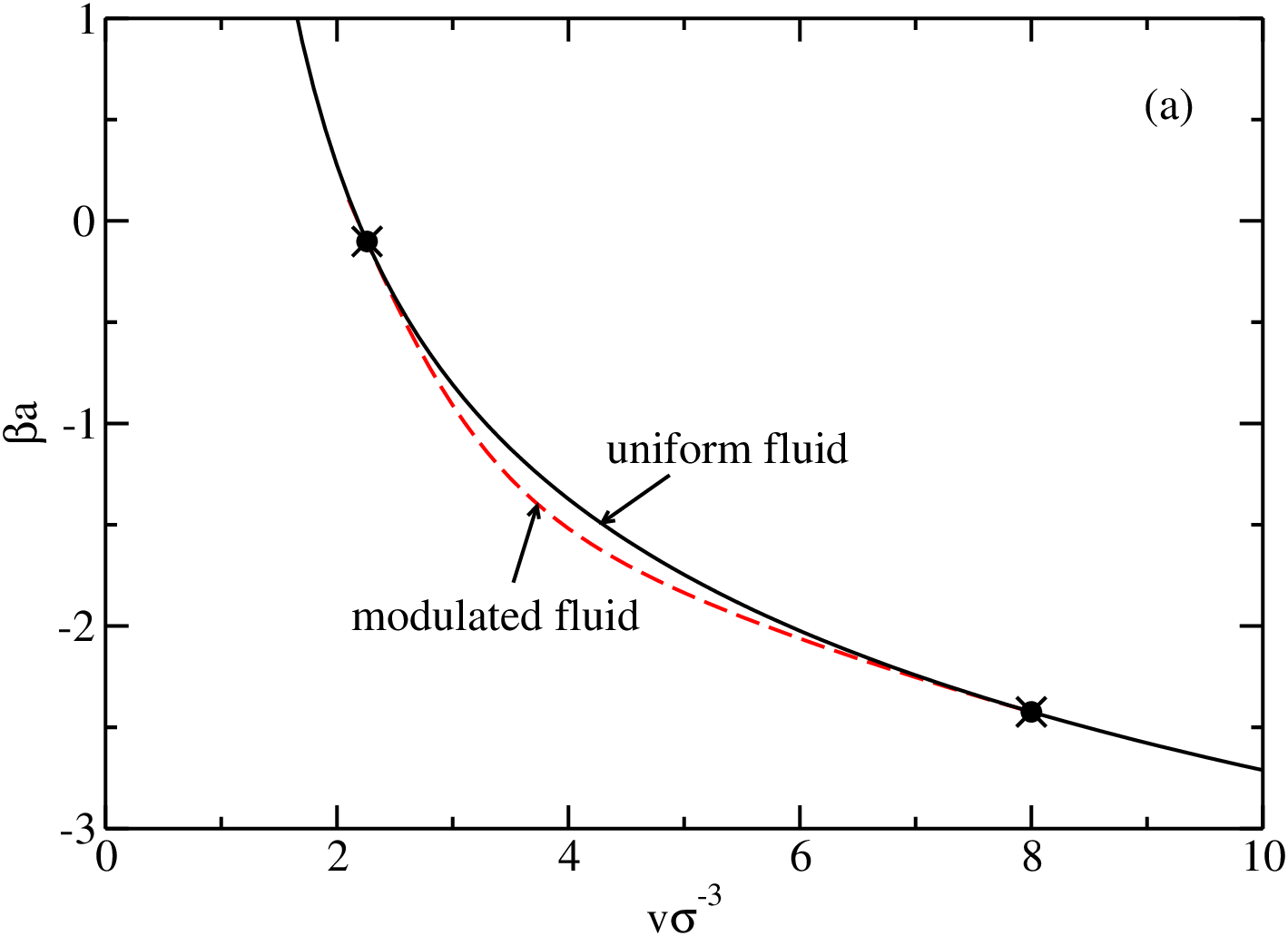}

\vspace{1.3cm}
\includegraphics[width=1.\columnwidth]{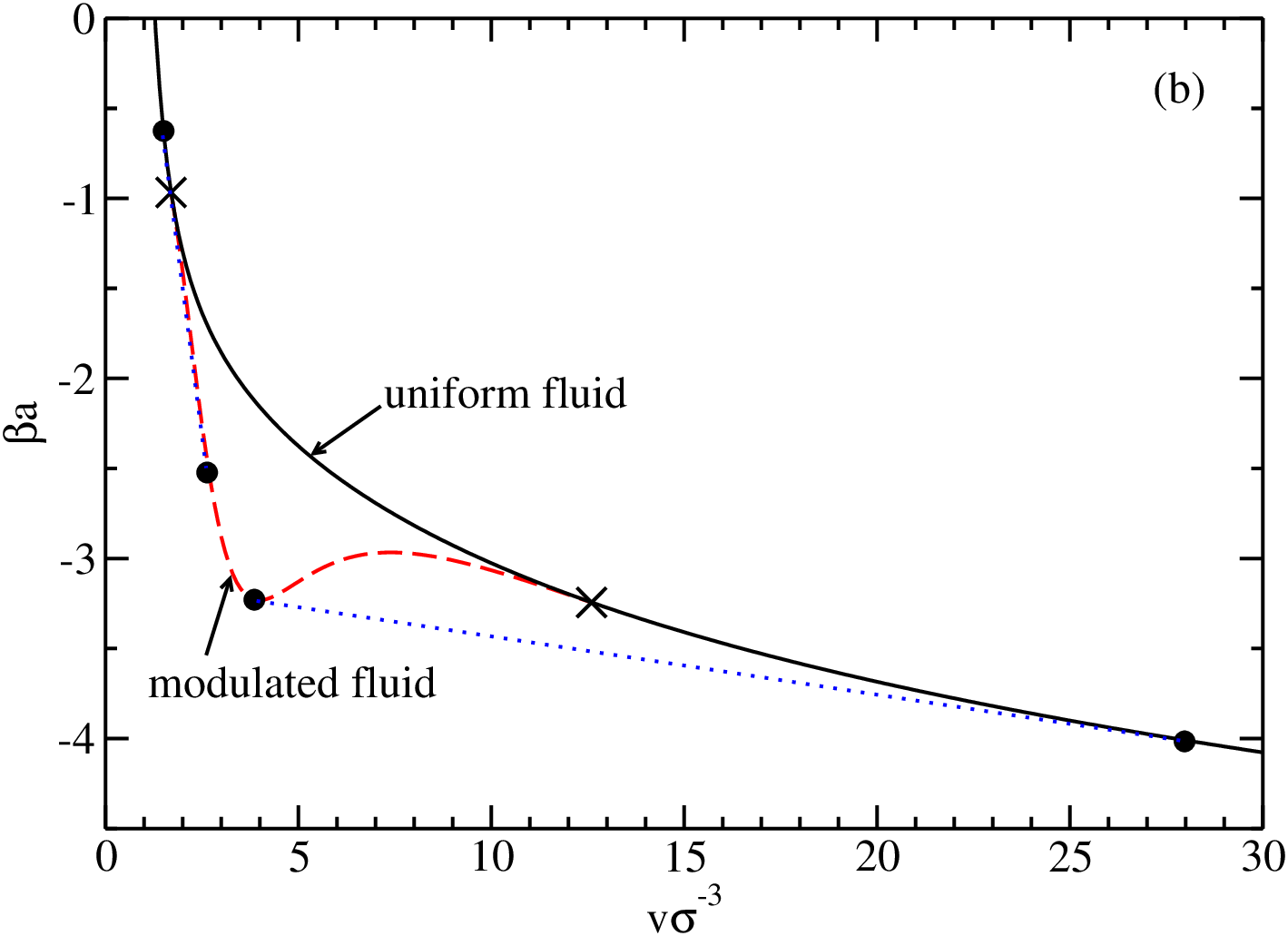}
\caption{The Helmholtz free energy per particle $a$ plotted as a function of $v$, the volume per particle, for the set of fluid pair potential parameters $A=0.5$, $\lambda_1=1$ and $\lambda_2=0.5$. The solid line is the free energy for the uniform fluid, given by Eq.\ \eqref{eq:F_bulk} and the dashed line is the free energy for the modulated fluid, given by Eq.\ \eqref{eq:Landau_sub}. Fig.\ (a) is for the case when $\epsilon^{-1}=0.83$, above the tricritical point, and Fig.\ (b) is when $\epsilon^{-1}=0.72$, below the tricritical point. The dotted lines show the common tangent construction between the coexisting state points, which are marked with the symbol $\bullet$. The points on the $\lambda$-line, where the two free energies are equal are marked with the symbol $\times$.
\label{fig:free_energy}}
\end{figure}

On the $\lambda$-line itself, $D(k) \geq 0$ for all values of $k$, with the equality being the case only for $k=k_c$. However, inside the $\lambda$-line, $D(k_c)<0$. Thus the coefficient of ${\cal A}^2$ in the free energy \eqref{eq:Landau} is negative for $k=k_c$, meaning that inside the $\lambda$-line the free energy is lower for some ${\cal A}>0$, than when ${\cal A}=0$, indicating that the modulated fluid has a lower free energy than the uniform fluid. If we neglect the terms ${\cal O}({\cal A}^6)$ and higher in the expansion \eqref{eq:Landau}, then from Eq.\ \eqref{eq:min_cond_2} we find that the minimum of the free energy is when the amplitude
\begin{eqnarray}
{\cal A}=\left(\frac{8|D(k_c)|}{\bar{\rho}B}\right)^{1/2}.
\label{eq:A_inside}
\end{eqnarray}
Substituting this value of ${\cal A}$ into Eq.\ \eqref{eq:Landau} we obtain
\begin{eqnarray}
\frac{\beta {\cal F}}{V}=\frac{\beta {\cal F}[\bar{\rho}]}{V}-\frac{D(k_c)^2}{\bar{\rho}^2B}
\label{eq:Landau_sub}
\end{eqnarray}
where we have neglected the contribution from terms of ${\cal O}({\cal A}^6)$ and higher in Eq.\ \eqref{eq:Landau_sub}. Since we have fairly simple expressions for all the terms in Eq.\ \eqref{eq:Landau_sub} [see Eqs.\ \eqref{eq:w_hat}, \eqref{eq:D}, \eqref{eq:lambda_line_2} and \eqref{eq:F_bulk}], we now have a simple expression for the Helmholtz free energy of the modulated phase.

Before using Eqs.\ \eqref{eq:F_bulk} and \eqref{eq:Landau_sub} for the free energy to examine the nature of the phase transition between the uniform and modulated fluid phases, we first recall that for two phases to coexist, the temperature, pressure and chemical potential in the two phases must be equal. As we show in the Appendix, these conditions correspond geometrically to making a common tangent construction on the free energy per particle $a(\bar{v})={\cal F}/N$ plotted as a function of $\bar{v}=V/N=1/\bar{\rho}$, the volume per particle ($N$ is the total number of particles in the system). In Fig.\ \ref{fig:free_energy} we display the free energy $a$ as a function of $\bar{v}$ for two different values of $\epsilon^{-1}$. In Fig.\ \ref{fig:free_energy}(a) we display the free energy for the set of pair potential parameters $\lambda_1=1$, $\lambda_2=0.5$, $A=0.5$ and $\epsilon^{-1}=0.83$. This curve is typical of the case when $\epsilon^{-1}>\epsilon_T^{-1}$, where $\epsilon^{-1}_T$ is the value of $\epsilon^{-1}$ at the tricritical point. In this case we see that there is no common-tangent construction -- i.e.\ the free energy is a convex function. Thus the phase transition between the uniform fluid and the modulated fluid occurs at the $\lambda$-line and the transition is second order. In fact, on increasing the density $\bar{\rho}$ (decreasing $\bar{v}$), we find there are two second order transitions. These two points are marked by the symbol $\bullet$ in Fig.\ \ref{fig:free_energy}(a), which are also two points on the $\lambda$-line. The low density tricritical point is at $\epsilon^{-1}=\epsilon^{-1}_T\simeq0.75$ and the higher density one is at $\epsilon^{-1}_T \simeq 0.79$.

On decreasing $\epsilon^{-1}$ below $\epsilon^{-1}_T$, one finds that the free energy is no longer convex and that the common tangent construction between the uniform fluid free energy and modulated fluid free energy can be made -- see for example the results for $\epsilon^{-1}=0.72$, displayed in Fig.\ \ref{fig:free_energy}(b). The common tangent construction lines are the two dotted lines. These lines join coexisting state points which are marked by the symbol $\bullet$ in Fig.\ \ref{fig:free_energy}(b) which lie at points either side of the $\lambda$-line (marked $\times$). The value of $\epsilon^{-1}$ at which the free energy curve goes from being convex, to non-convex defines $\epsilon_T^{-1}$ -- this is the tricritical point \cite{ChaikinLubensky}.

\begin{figure}
\includegraphics[width=1.\columnwidth]{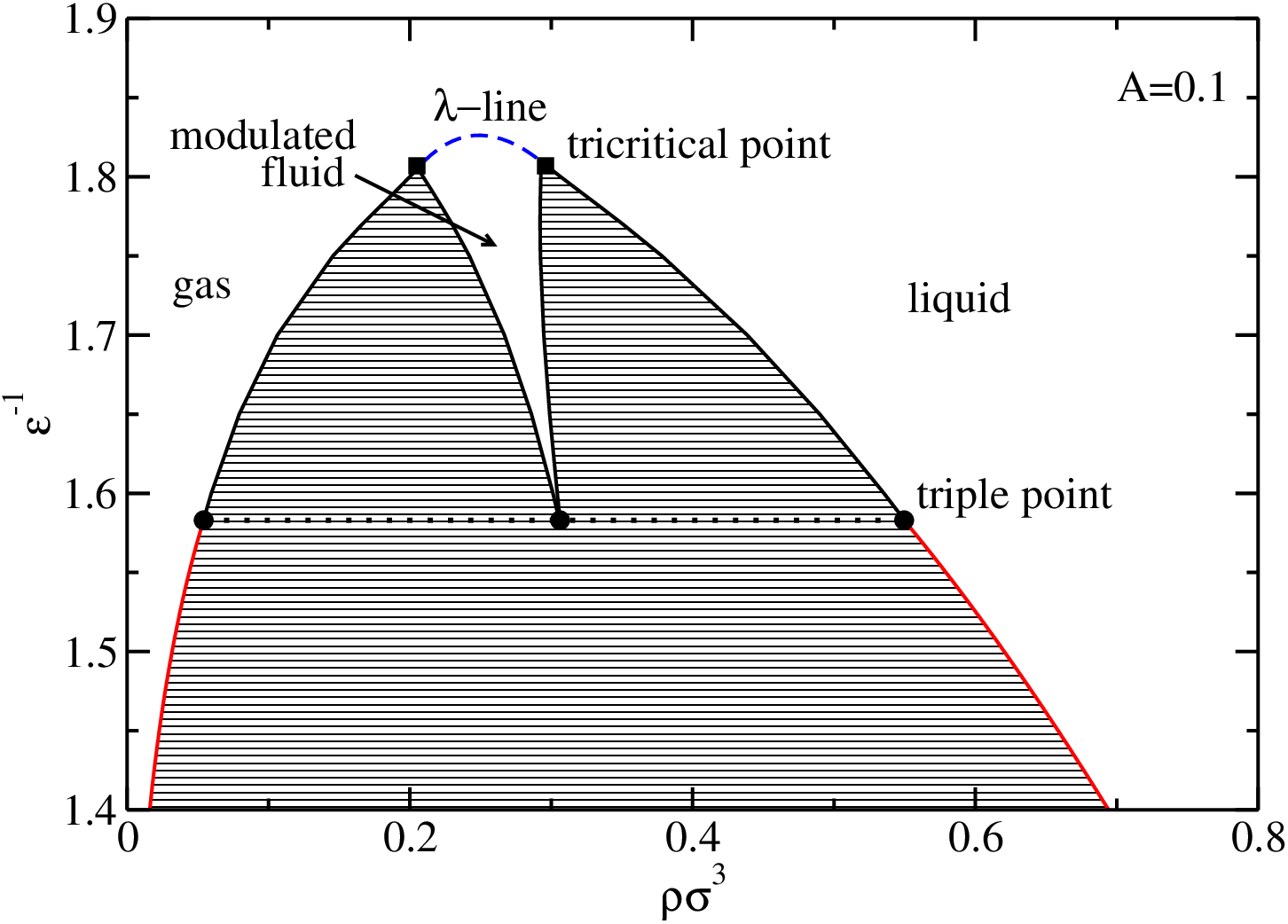}

\vspace{1.3cm}
\includegraphics[width=1.\columnwidth]{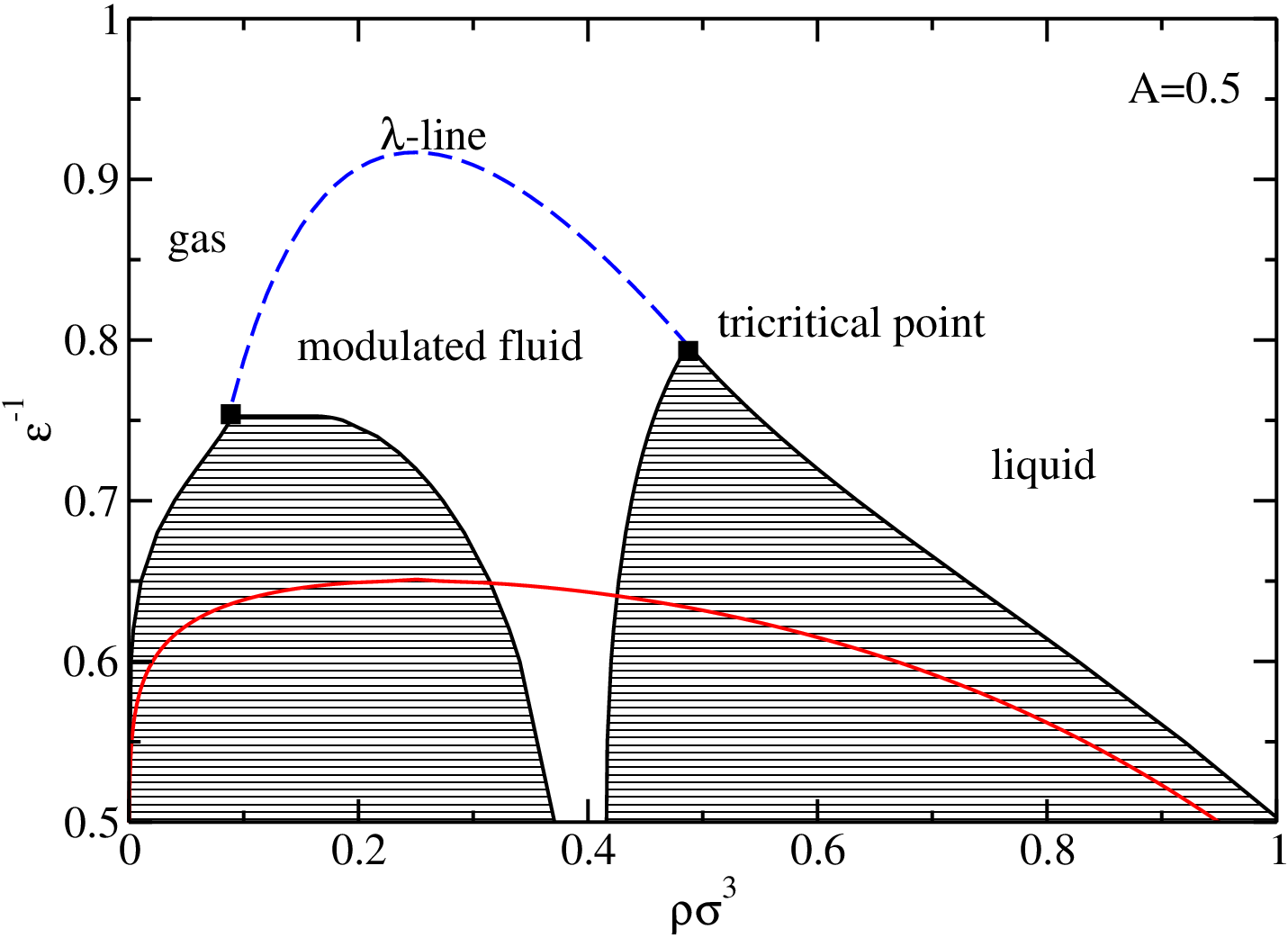}
\caption{Phase diagram in the $\epsilon^{-1}$ versus density $\rho$ plane, for the system with pair potential parameters $\lambda_1=1$, $\lambda_2=0.5$ and (in the top diagram) $A=0.1$ or (in the bottom diagram) $A=0.5$, obtained using Eq.\ \eqref{eq:Landau} truncated after the term of ${\cal O}({\cal A}^6)$.
\label{fig:phase_diag}}
\end{figure}

In Fig.\ \ref{fig:phase_diag} we display the phase diagram resulting from performing the common tangent construction on the free energy for a range of values of $\epsilon^{-1}$. We also display the phase diagram when the pair potential parameter $A=0.1$. In calculating these phase diagrams we included the term of ${\cal O}({\cal A}^6)$ in the free energy, since truncating the expansion of the free energy in powers of the order parameter ${\cal A}$ is only really justified when ${\cal A}$ is small. Even including the term of ${\cal O}({\cal A}^6)$, one should not expect the theory to be reliable for determining the coexistence curve between the uniform fluid and the modulated fluid for $\epsilon^{-1}<\epsilon^{-1}_T$, away from the tricritical point. Owing to this, the phase diagrams in Fig.\ \ref{fig:phase_diag} are only qualitatively correct for $\epsilon^{-1}<\epsilon^{-1}_T$, where the truncated theory predicts the free energy of the modulated phase to be lower than it really is.
For the case $A=0.5$ the theory predicts (incorrectly) that as $\epsilon^{-1}$ is decreased, the two first order transition lines never intersect, as they do for $A=0.1$.
This is the reason why the theory (incorrectly) predicts that there is no triple point for the case when $A=0.5$.

A final question we wish to address in this section is: How does the amplitude ${\cal A}$ grow as a function of distance in the phase diagram from the $\lambda$-line? Let us denote the fluid density on a the $\lambda$-line itself, for a certain value of $\epsilon^{-1}>\epsilon^{-1}_T$, as $\rho_{\lambda}$. Recall that $D(k_c,\rho_\lambda)=0$, which means that $D(k_c) \sim (\bar{\rho}-\rho_\lambda)$ near the $\lambda$-line. Combining this with Eq. \eqref{eq:A_inside}, we find that inside and near to the $\lambda$-line the amplitude 
\begin{equation}
{\cal A} \sim (\bar{\rho}-\rho_\lambda)^{1/2}
\label{eq:A}
\end{equation}
and, of course, outside the $\lambda$-line the amplitude ${\cal A}=0$.

\section{DFT results}
\label{sec:DFT_res}

\begin{figure}
\includegraphics[width=1.\columnwidth]{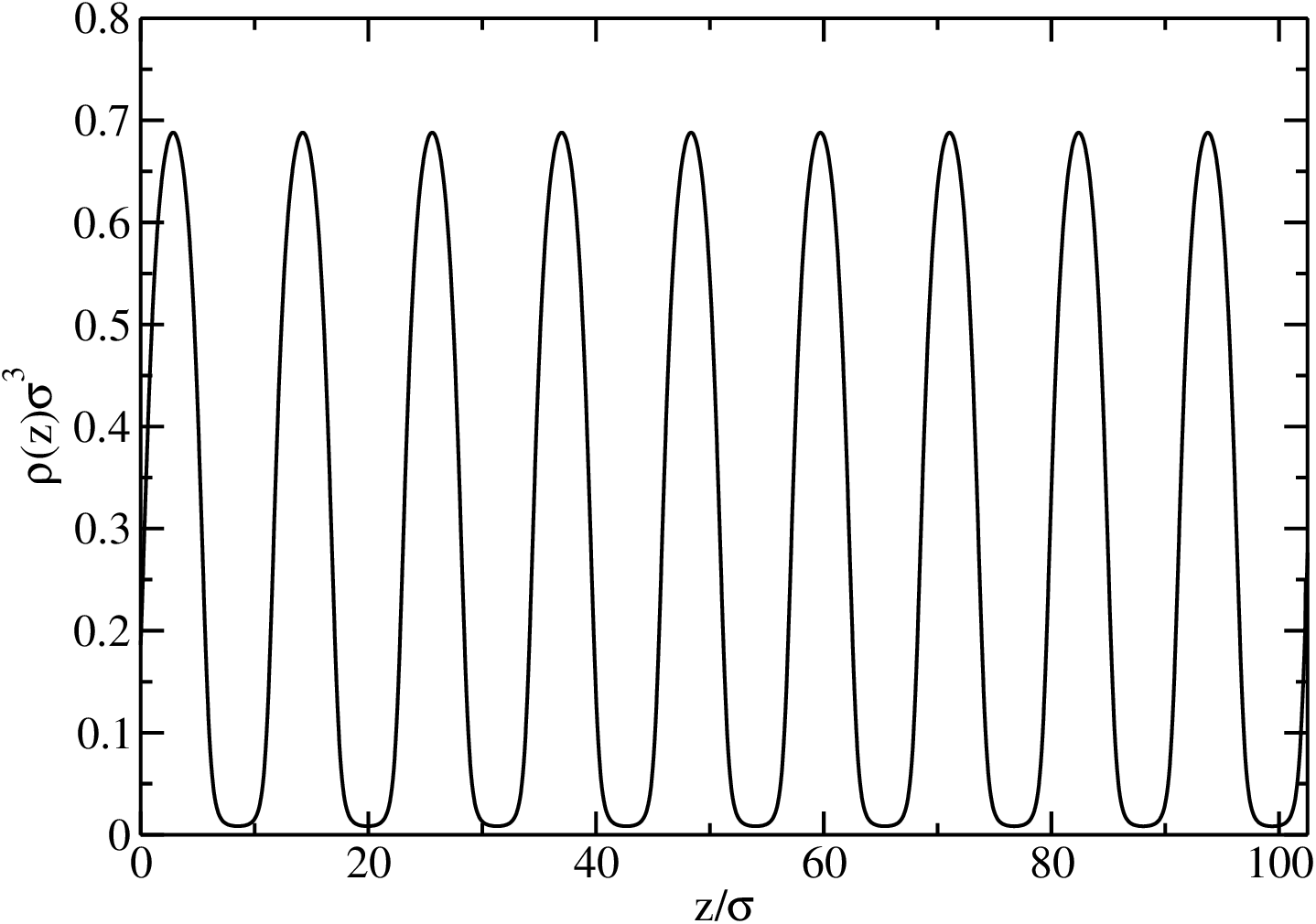}
\caption{Equilibrium density profile inside the modulated phase region, for $\epsilon^{-1}=0.65$ and $\bar{\rho}\sigma^3=0.29$. The same choice of the potential range and strength was made calculating the phase diagram in Fig.\ \ref{lambdafull}, where we see that this density profile corresponds to a point well inside the modulated phase region.
\label{amplitudeoscil}}
\end{figure}

\begin{figure}
\vspace{1cm}
\includegraphics[width=1.\columnwidth]{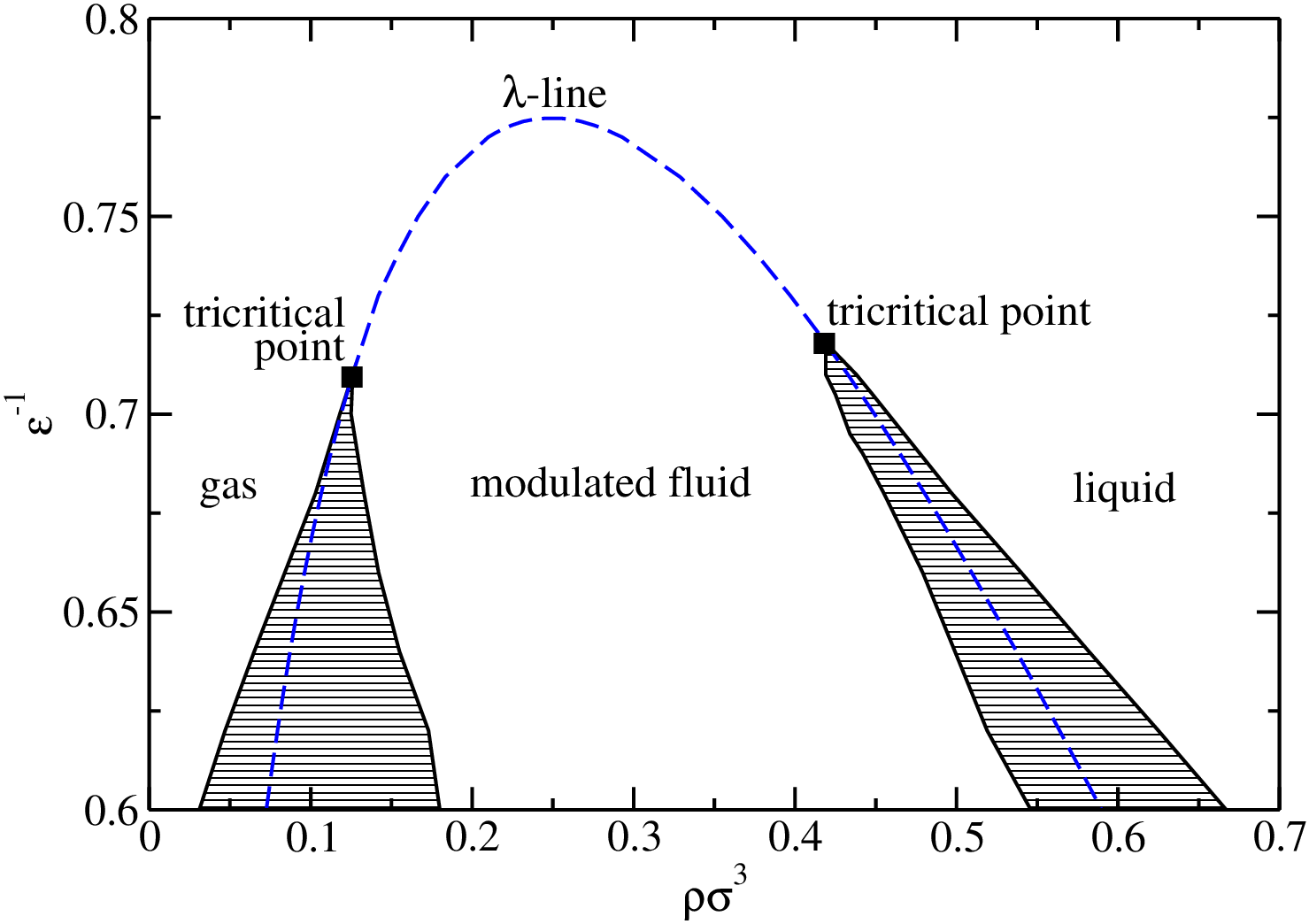}
\caption{Phase diagram in the $\epsilon^{-1}-\rho$ plane obtained from the RPA DFT for the parameters $ \lambda_1 =1,  \lambda_2 =0.5$ and $A=0.5$. The shaded regions are the two phase coexistence regions between the uniform and modulated fluid phases. They end at the tricritical points, where the transition ceases to be first order and becomes second order. Above the tricritical points, the transition line is the $\lambda$-line.
\label{lambdafull}}
\end{figure}

In this section we present results from numerically solving the full DFT theory \eqref{eq:F}, by finding the solution to the Euler-Lagrange equation \eqref{miusolve}, without resorting to the sinusoidal approximation that was used to develop the order parameter theory in the previous section.

To calculate the density profile that minimises the grand potential functional for a given point in the phase diagram, i.e.\ for given values of $\epsilon^{-1}$ and chemical potential $\mu$, we solve the Euler-Lagrange equation \eqref{miusolve} using a simple iterative scheme. To determine the equilibrium density profile for the bulk system, i.e.\ when $V_{ext}(z)=0$, we solve for the profile using periodic boundary conditions. Note that in this case a uniform (constant) density profile $\rho(z)=\bar{\rho}$ corresponds to a stationary curve for the functional \eqref{eq:F}. In order to find the minimum of the free energy corresponding to an oscillatory density profile, such as that displayed in Fig.\ \ref{amplitudeoscil}, we must break the symmetry by using a non-uniform initial guess for the density profile in our iterative solver. The final density profile does not depend on the precise form of the initial guess; we used either a step function or the density profile from a neighbouring state point as our initial guess in this study. We also found it useful to fix the value of the density profile $\rho(z)$ to be $\bar{\rho}$ at a single point. This was done for the following reason: If a particular density profile $\rho(z)$, such as that displayed in Fig.\ \ref{amplitudeoscil}, is the equilibrium density profile, then so is $\rho(z+l)$, where $l$ is any real number. Thus, any numerical optimisation algorithm will initially descend to a minimum of the free energy corresponding (say) to the density profile $\rho(z)$. However, due to numerical errors the gradient of the free energy landscape will not {\em exactly} be zero for this profile and so the optimisation routine will then perform a very slow walk along the free energy valley, where points along the bottom of the valley correspond to different values of $l$. By fixing the density at a single point, one penalises such translations.

A further issue to consider is that since we solve for $\rho(z)$ on a finite grid of length $L$ with periodic boundary conditions, this length $L$ must be commensurate with the periodicity of the modulations in the density profile. Thus we must also minimise the free energy with respect to variations in $L$, or, equivalently, with respect to variations in the spacing between density grid points.

Phase coexistence between the uniform fluid and the modulated fluid phases is determined by calculating the grand potential $\Omega$ for fixed values of $\epsilon^{-1}$ (which is equivalent to fixing the temperature), whilst slowly varying the chemical potential $\mu$. Recall that two phases coexist if the chemical potential, pressure (recall $\Omega=-PV$, where $P$ is the pressure) and temperatures are equal in the two phases. We display the resulting phase diagram in Fig.\ \ref{lambdafull}. The shaded region denotes the two phase coexistence region. This phase diagram should be compared with the lower phase diagram in Fig.\ \ref{fig:phase_diag}, which was obtained using the order parameter theory of the previous section (recall, however, that Fig.\ \ref{fig:phase_diag} was obtained using a different expression for $w(r)$ {\em within} the hard-sphere core $r<\sigma$, than the DFT used to obtain the results in Fig.\ \ref{lambdafull} -- see the discussion in Sec.\ \ref{sec:DFT}). The boundaries of the two phase region lie either side of the $\lambda$-line. As the value of $\epsilon^{-1}$ is increased these coexistence lines meet at the $\lambda$-line. This meeting point is a tricritical point. Above the tricritical points, the phase transition is second order and follows the $\lambda$-line exactly.

\begin{figure}
\includegraphics[width=1.\columnwidth]{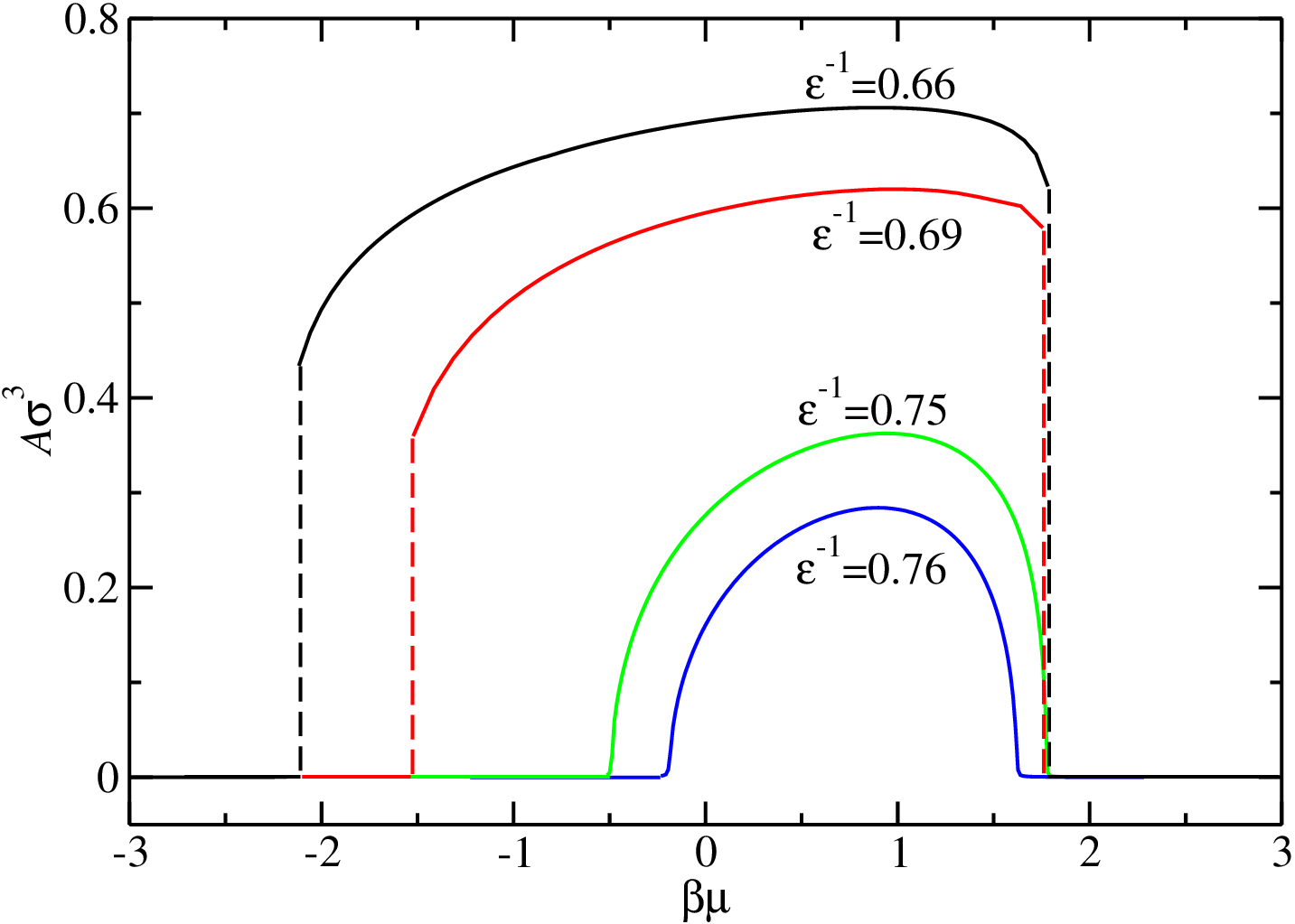}
\caption{Variation of the order parameter, the  amplitude ${\cal A}$, as a function of the chemical potential $\mu$, for various values of $\epsilon^{-1}$. The curves for $\epsilon^{-1}=0.75$ and $0.76$ are both continuous and correspond to values of $\epsilon^{-1}$ above the tricritical points. The curves for $\epsilon^{-1}=0.66$ and $0.69$ have discontinuities (marked by the dashed lines) at the phase transitions and these curves correspond to values of $\epsilon^{-1}$ below the tricritical points.
\label{fig:A_vs_mu}}
\end{figure}

\begin{figure}
\includegraphics[width=1.\columnwidth]{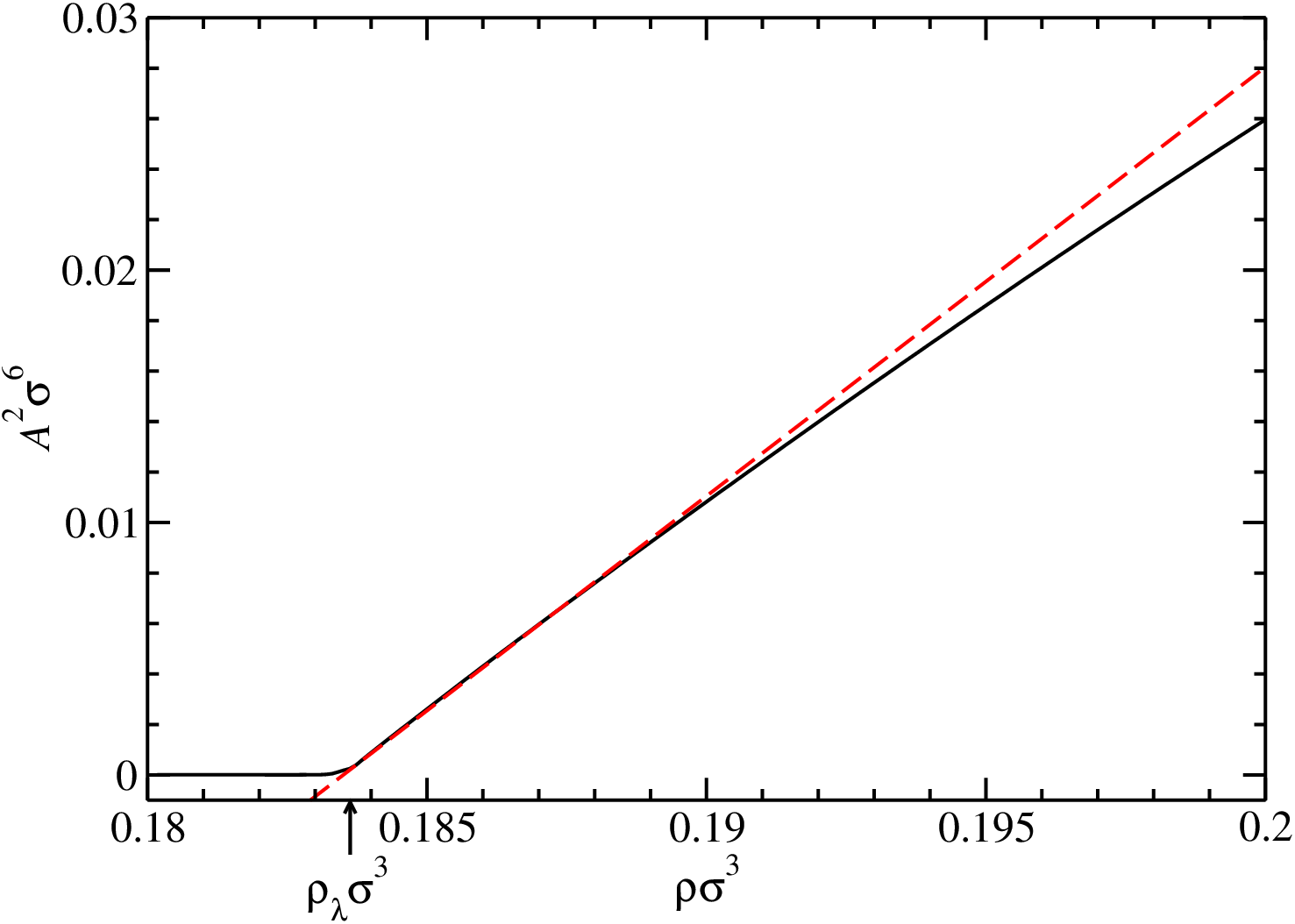}
\caption{Plot of the amplitude squared ${\cal A}^2$ versus density $\bar{\rho}$ for the parameters $A=0.5$ and $\epsilon^{-1}=0.76$. The straight line is the line ${\cal A}^2=1.7(\bar{\rho}-\rho_\lambda)$, where $\rho_\lambda\sigma^3=0.1835$.
\label{fig:A_squared}}
\end{figure}

In Fig.\ \ref{fig:A_vs_mu} we display plots of the amplitude as a function of $\mu$ for a number of different values of $\epsilon^{-1}$. For $\epsilon^{-1}<\epsilon^{-1}_T$, we see that the amplitude varies discontinuously at the phase transition, jumping from ${\cal A}=0$ in the uniform phases to a non-zero value in the modulated phase. However, for $\epsilon^{-1}>\epsilon^{-1}_T$ we see that the amplitude changes continuously across the phase transition. In Fig.\ \ref{fig:A_squared} we plot the amplitude squared as a function of $\bar{\rho}$ for case when $\epsilon^{-1}=0.76>\epsilon^{-1}_T$. We see that when ${\cal A}$ is small, ${\cal A}^2 \propto (\bar{\rho}-\rho_\lambda)$, indicating that the result in Eq.\ \eqref{eq:A}, obtained from the order parameter theory, also applies to solutions of the full DFT.

\section{Discussion and Conclusions}
\label{sec:conc}

In this paper we have used DFT to study systems composed of particles with competing interactions. This model fluid exhibits phase transitions from the uniform fluid to a modulated fluid phase. In order to elucidate the precise nature of the phase behaviour of this system we developed an order parameter (Landau) theory for the phase transition, using the amplitude ${\cal A}$ of the modulations as the order parameter. Both the full DFT theory and the order parameter theory predict the following phase behaviour: when the parameter $\epsilon^{-1}<\epsilon_T^{-1}$, on increasing the fluid density one finds that there is a first order phase transition from the uniform fluid phase to the modulated phase. At the transition point, there is discontinuous change in both density $\bar{\rho}$ and the amplitude ${\cal A}$, which jumps from ${\cal A}=0$ to ${\cal A} \neq 0$ discontinuously. However, when $\epsilon^{-1}>\epsilon_T^{-1}$, on increasing $\bar{\rho}$ one finds that there is a second order (continuous) phase transition from the uniform fluid phase to the modulated phase. At the transition point, which is the $\lambda$-line, both density $\bar{\rho}$ and the amplitude ${\cal A}$ vary continuously. Both outside and on the $\lambda$-line itself, ${\cal A}=0$. On moving off the $\lambda$-line, one finds that the amplitude ${\cal A}$ increases in a continuous manner. The precise form of this increase, for small values of ${\cal A}$, is given by Eq.\ \eqref{eq:A}.

The phase behaviour predicted by the theory in this paper is in qualitative agreement with the results observed recently in Monte Carlo computer simulations of the present model \cite{Archer21}. However, one should bear in mind that the present theory is a mean field theory. The phase behaviour of the present system is somewhat analogous to that observed in diblock copolymer systems \cite{BarratHansen, LeiblerMacrom1980, FredricksonHelfandJCP1987}.  In these systems, the result is that when fluctuations beyond mean field are taken into account, the second order transition becomes weakly first order \cite{FredricksonHelfandJCP1987}. This scenario is not supported for the present system by the Monte Carlo simulations in Ref.\ \cite{Archer21}, but it may be the case that for some choices of parameters not explored in Ref.\ \cite{Archer21}, that the transition at the $\lambda$-line becomes weakly first order.

We should also remind the reader that in the present study we have assumed throughout that the density profile varied only in one spatial dimension. Whilst this is true for any lamellar phase, one should expect the present system to exhibit cluster, bubble and perhaps other modulated phases \cite{Archer21}. The density profile for these phase will vary in more than one Cartesian direction. We plan to study the phases exhibited by the present DFT when the theory is not constrained to exhibiting modulations in only one direction. We expect the region in phase diagrams \ref{fig:phase_diag} and \ref{lambdafull} labelled `modulated fluid' to be further subdivided into a number of different modulated phases, in a manner somewhat analogous to the modulated phases displayed by a two dimensional fluid with competing interactions \cite{ImperioReattoJPCM2004, ImperioReattoJCP2006}.

The technological applications of fluids exhibiting modulated phases could be significant. For example, in display technologies or in making masks for micro-lithography. It is therefore important to understand and control the formation of the various type of modulated structures exhibited by these systems. The present study goes some way towards this goal.

\section*{Acknowledgements}

A.J.A. is grateful for the support of RCUK and C. I., D. P., and L. R. acknowledge support from the European Union, Contract No. MRTN-CT2003-504712. We are grateful to Bob Evans and Nigel Wilding for several helpful discussions.

\appendix*
\section{Common tangent construction}

In this Appendix we show that the common tangent construction on the Helmholtz free energy per particle $a={\cal F}/N$ yields the coexisting state points.

From the free energy we may obtain the following quantities: (i) the pressure
\begin{eqnarray}
P=\left(\frac{\partial {\cal F}}{\partial V} \right)_{N,T}=-\left( \frac{\partial a}{\partial \bar{v}}\right),
\end{eqnarray}
where $\bar{v}=V/N=1/\bar{\rho}$ is the volume per particle and (ii) the chemical potential
\begin{eqnarray}
\mu=\left(\frac{\partial {\cal F}}{\partial N} \right)_{V,T} = a-\bar{v} \left( \frac{\partial a}{\partial \bar{v}}\right).
\end{eqnarray}
If $\bar{v}_1$ and $\bar{v}_2$ are the coexisting volumes per particle, then the conditions for mechanical equilibrium $P(\bar{v}_1)=P(\bar{v}_2)$ and of chemical equilibrium $\mu(\bar{v}_1)=\mu(\bar{v}_2)$ give us
\begin{eqnarray}
\frac{\partial a}{\partial \bar{v}} \Bigg\vert_{\bar{v}_1}
=\frac{\partial a}{\partial \bar{v}} \Bigg\vert_{\bar{v}_2}
=\frac{a(\bar{v}_1)-a(\bar{v}_2)}{\bar{v}_1-\bar{v}_2}
\end{eqnarray}
which geometrically corresponds to the common tangent construction on $a(\bar{v})$.

\end{document}